\documentclass[aps,twocolumn]{revtex4-1}
\usepackage{graphicx}
\usepackage[justification=justified,width=\linewidth]{caption}
\usepackage{subcaption}
\usepackage{amsmath}
\usepackage{amsfonts}
\usepackage{amsthm}
\usepackage{amssymb}
\usepackage{amsbsy}
\usepackage{wasysym}
\usepackage{bm}
\usepackage{mathrsfs}
\usepackage{color}
\usepackage{times}
\usepackage[resetlabels]{multibib}
\begin{document}
	
	\title{Surface directed spinodal decomposition of fluids confined in cylindrical pore}
	\author {Daniya Davis and Bhaskar Sen Gupta}
	\email{bhaskar.sengupta@vit.ac.in}
	\affiliation{Department of Physics, School of Advanced Sciences, Vellore Institute of Technology, Vellore, Tamil Nadu - 632014, India}
	
	\date{\today}

\begin{abstract}

The surface directed spinodal decomposition of a binary liquid confined inside cylindrical pore is investigated using molecular dynamics simulation. One component of the liquid wets the pore surface while the other remains neutral. A variety of wetting conditions are studied. For the partial wetting case, after an initial period of phase separation, the domains organize themselves into plug-like structure and the system enters into a metastable state. Therefore, a complete phase separation is never achieved. Analysis of domain growth and the structure factor suggests an one-dimensional growth dynamics for partial wetting case. As the wetting interaction is increased beyond a critical value, a transition from the plug-like to tube-like domain formation is observed which corresponds to the full wetting morphology. Thus, a complete phase separation is achieved as the wetting species moves towards the pore surface and forms layers enclosing the non wetting species residing around the axis of the cylinder. The coarsening dynamics of both the species are studied separately. The wetting species is found to follow a two-dimensional domain growth dynamics with a growth exponent $1/2$ in the viscous hydrodynamic regime. This was substantiated by the Porod tail of the structure factor. On the other hand, the domain grows linearly with time for the non wetting species. This suggests that the non wetting species behaves akin to a three-dimensional bulk system. An appropriate reasoning is presented to justify the given observations.

\end{abstract}

\maketitle

\section{Introduction}
The kinetics of phase separation of fluids in confinement are of high importance in scientific research~\cite{Brochard,De,Maher,Goh} as well as in the industry~\cite{Kanamori}. There are boundless applications of phase separating fluids in confinement. Especially the oil, gasoline and natural gas extraction industries are highly relied on these phenomena~\cite{Morrow}. Nonetheless, many possibilities are still unexplored and  plenty of questions regarding phase separation in such systems are unanswered. In this context, there is a paramount importance to study the transformation of single as well as multi-component phase separating fluid mixtures.

When a homogeneous binary liquid system is rapidly cooled within the miscibility gap, it loses thermodynamic stability and undergoes phase separation, forming distinct regions or domains. Over time, these domains grow and evolve until a state of local equilibrium or saturation is reached~\cite{Binder,Siggia,Furukawa,Miguel,Tanaka,Beysens,Rounak1,Rounak3,Puri,Dutt,Laradji,Thakre,Ahmad}. However, a system under confinement behaves differently from its bulk counterpart due to the presence of additional factors like restriction, surface effects, and system size. For instance, under confinement, the emergence of anisotropic domain growth becomes apparent. This phenomenon primarily arises from the constraints imposed by the limited capacity of particles that can occupy a confined space. The salient features of phase separation in such systems are metastability and lack of observable macroscopic phase separation. In real experiments, physical systems are often enclosed within containers or possess exposed surfaces, which typically exhibits a preferential attraction towards one of the species of the mixture. This selective affinity can significantly influence the rate of phase separation. This phenomenon, referred to as the wetting effect, entails a continual and persistent competition between phase separation and interactions with the surface or wall.

The coarsening process is always affected by the nature of the system. Usually a single time dependent length scale $\ell(t)$ characterizes the domain morphology~\cite{Bray}. This is obtained from the equal time correlation function $C(\vec r,t)$, where $\vec r$ is the distance between two spatial points and $t$ is the time after quench. The average domain size of the system follows the power law $\ell(t)  \sim t^{\alpha} $, where $\alpha$ is the growth exponent. The value of $\alpha$ is determined by corresponding coarsening mechanism that drives the phase separation. 

For the phase separation in solid-solid mixtures, diffusion takes precedence, and the growth exponent is $\alpha=1/3$~\cite{Chen}.
However, in fluid systems, the hydrodynamic effect becomes significant and the growth exponents change accordingly. In fluid-fluid mixtures the diffusive phase is short lived and the system quickly transits to hydrodynamic regime. Here, we have two exponents corresponding to the viscous hydrodynamics ($\alpha =1$) and inertial hydrodynamics ($\alpha =2/3$). The results mentioned above refer to the bulk systems~\cite{Siggia, Furukawa}.

For the phase separation of fluids in confined geometry, existing studies predominantly employ two primary methods of analysis. The first one involves utilization of the random pore Ising model~\cite{De,Goh,Wiltzius}, which maps the system onto a network of random pores. The second method, known as the single pore model~\cite{Liu} is a widely accepted model for studying phase separation of liquids inside porous media, and does not rely on mapping to any specific model or randomness. For the later, theoretical studies were conducted, focusing on the wetting behavior of a binary fluid system inside a cylindrical pore~\cite{Liu}. This study introduced the benchmark single pore model, which allowed to examine the phase separation in confined space, particularly applicable to scenarios such as binary fluid segregation within vycor glasses where the random pore Ising model is not suitable due to low-porosity. The transition of the liquid structure from a plug-like to a tube-like form was illustrated via a wetting phase diagram. In between, there exists an intermediate capsule-like structure, which occurred only when the radius of the pore is relatively larger. The domain growth was found to slow down when it became comparable to the pore size. 

The phase separation of binary liquid inside a two-dimensional porous media was studied by numerically integrating the Cahn-Hilliard equation with and without wetting effects~\cite{Chakrabarti}. While the random field Ising model failed to explained the slowing down of the domain growth and the breakdown of scaling laws in such systems, the single pore model successfully explained the source of slow growth. Subsequent work on binary liquid inside two-dimensional strip geometry involving the numerical study of Cahn-Hilliard-Cook equation~\cite{Rao} further confirmed the validity of the single pore model. Later on, this work was extended to study the effect of a variety of asymmetric pores, i.e. a simple strip pore, an uneven single pore, and a junction made out of two pores~\cite{Zhang}. The single pore method was explored further to study the liquid-liquid phase separation using molecular dynamics simulations with neutral pore wall (no wetting)~\cite{Gelb1,Gelb2}.

The surface directed spinodal decomposition in binary liquid mixture was studied in the bulk system using a mesoscopic-level modeling in terms of coupled Vlasov-Boltzmann equations with long range interactions~\cite{Bastea}. The effect of weak and strong surface field on the domain growth was analysed. A two-dimensional study on how the wetting effect towards the mobile and immobile particles in the binary fluid system affects the phase separation of the later was examined in ref.~\cite{Araki}. Similarly, surface field study was conducted numerically and attention was paid to obtain the standard growth laws in the bulk region of the system~\cite{Jaiswal}. Recently,  molecular dynamics simulation was carried out on the binary fluid inside the cylindrical nanopore with neutral wall and the growth nature of the domain was studied before the system attained a metastable state~\cite{Basu}. An early time diffusive growth was observed and the later time growth exponent was found to match with the inertial hydrodynamic growth in the two-dimensional bulk system.

However, the evolution of domains of segregating fluids inside the single pore cylindrical tube in the presence of wetting interaction of a preferred component of the liquid with the confining wall has not been addressed properly till now. In particular, the effect on the domain structures and the growth laws when the wetting interaction is systematically changed is missing.  As previously mentioned, this model becomes more representative of experimental observations when the influence of wetting effects is taken into account. In this paper we use extensive molecular dynamic simulation to study the kinetics of persistent interplay between the phase separation and wetting, deep within the coexistence curve. 

\section{Models and methods}
In this study we use a binary AB liquid mixture confined inside a cylindrical pore using molecular dynamic simulation. The fluid particles interact with each other via the Lennard Jones (LJ) potential
\begin{equation}
 \label{eq:lj_potential}
	U_{\alpha\beta}(r)=4\epsilon_{\alpha\beta}\left[\left(\frac{\sigma_{\alpha\beta}}{ r_{ij}}\right)^{12}- \left(\frac{\sigma_{\alpha\beta}}{r_{ij}}\right)^6\right]
\end{equation}
 where $\epsilon$ is the interaction strength, $\sigma$ is the particle diameter, $r_{ij}=|r_i-r_j|$ is the scalar distance between the two particles $i$ and $j$ and $\alpha$, $\beta$ $\in$ A, B. The phase separation between the two types of particles is assured by assigning the interparticle diameters as $\sigma_{AA}=\sigma_{BB}=\sigma_{AB}=1.0$ and the interaction parameters as $\epsilon_{AA}=\epsilon_{BB}=2\epsilon_{AB}=\epsilon$. This method can be mapped to Ising model. The computational load is reduced by assigning a cut-off at $r=r_c=2.5$ for the LJ potential.  This cut-off introduces a discontinuity in the potential and force term. This is resolved by modifying the potential as 
 \begin{equation}
\label{eq:mod_potential}
u(r)=U(r)-U(r_c)-(r-r_c)\left(\frac{dU}{dr}\right)|_{r=r_c} 
\end{equation}
The final term in the Eq.~\ref{eq:mod_potential} avoids the abrupt jumps in the force at $r_c$. The system mentioned above is characterized in bulk with a critical temperature of $T_c=1.421$ and critical density of $\rho_c=N/V=1$ in three-dimension~\cite{Das}. Here $N$ is the number of particles in the system and $V$ is the volume. We measure the temperature and length in units of $\epsilon/k_B$ and $\sigma$ respectively. For convenience $\epsilon$, $k_B$ and mass of each particle $m$ are set to unity.

A cylindrical tube with a large length to diameter ratio is considered, which serves as a confining structure containing the binary mixture. The axis of the cylinder is chosen to be the x-axis. A periodic boundary condition is applied along the length of the cylinder. The wall of the cylinder is constructed with closely packed particles similar to that of the fluid particles. The wetting effect is incorporated by introducing a preferable attraction of one type of particles, say type A towards the wall via the LJ potential given in Eq.~\ref{eq:mod_potential}, and no interaction at all for the other species. The interaction between the wall and type A particles, denoted as $\epsilon_w$ is tuned over a wide range of values and the effect of this wetting strength on the phase separation is studied. We vary the $\epsilon_w$ in the range of (0.1, 0.8).

Molecular dynamic simulation (MD) is performed in the canonical ensemble. Since our system is in liquid state, it is important to take into consideration the effect of hydrodynamics. Therefore, Nose Hoover thermostat is used which controls the temperature, and at the same time preserves the hydrodynamics of the system~\cite{Nose}. Velocity-Verlet Algorithm is used in the MD simulation to compute the positions and velocities of the particles with timestep of  $\Delta t=0.005$~\cite{Verlet}. Here time is measured in units of $(m\sigma^2/\epsilon)^{1/2}$. 

The cylindrical pore we consider has a radius $R=10$ and length $L=200$. It is filled with the binary liquid of number density $\rho=0.8$, where $50\%$ of the particles are type A and $50\%$ type B. The system is first equilibrated at a high temperature of $T_i=10$ to prepare a homogeneous mixture and then suddenly quenched to a temperature $T_f=0.8$ well below $T_c$. Finally, the time evolution of the system towards the thermodynamically favored state at $T_f=0.8$ is studied. The results are averaged over 80 independent initial configurations.   

To study the domain growth and coarsening dynamics of the segregating liquid inside the cylindrical pore, we use the so-called two-point equal time correlation function $C(\vec r,t)$ given by
\begin{equation}
    \label{eq:correlation}
     C(\vec r,t)=\langle\psi(0,t)\psi(\vec r,t)\rangle-\langle\psi(0,t)\rangle\langle\psi(\vec r,t)\rangle.
 \end{equation}
The angular brackets represent the ensemble averaging. $\psi(\vec r,t)$ is the order parameter of the system defined in terms of the local density fluctuations as 
\begin{equation}
 	\label{eq:order_parameter}
    \psi(\vec r,t)={\frac{\rho_A(\vec r,t)-\rho_B(\vec r,t)}{\rho_A(\vec r,t)+\rho_B(\vec r,t)} }
\end{equation} 
Here $\rho_A(\vec r,t)$ and $\rho_B(\vec r,t)$ are the local concentrations of A and B particles at time t around the position $\vec r$. For the domain structure related studies, we resort to the static structure factor $S(\vec k,t)$, obtained from the Fourier transformation of the correlation function given by
 \begin{equation}
 	\label{eq:structure factor}
   S(\vec k,t)=  \int d\vec r\ exp(i\vec k.\vec r)\ C(\vec r,t) 
 \end{equation}
 where $k$ is the wave vector~\cite{Bray}. For the large-$k$ limit in $d$-dimension, the $S(\vec k,t)$ follows the Porod law given by
  \begin{equation}
 	\label{eq:porodLaw}
 	S(k,t) \sim k^{(-d+1)} 
 \end{equation}
 A detailed description of computing the order parameter $\psi(\vec r,t)$ under different wetting conditions is provided in the next section. 
 
\section{Results}

\begin{figure}[h]
	\centering
	\includegraphics[width=0.95\columnwidth]{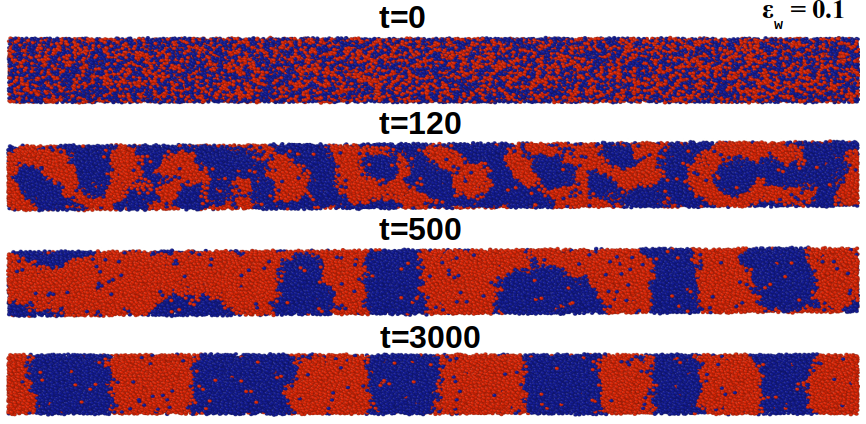}
	\caption{Time evolution of the segregating binary liquid mixture confined inside cylindrical pore for the partial wetting interaction $\epsilon_w=0.1$.  A and B type of particles are represented by red and blue colors respectively.}
	\label{fig:ea01}
\end{figure}

It is well-established that in a bulk system, when our symmetric binary liquid is quenched below the critical temperature, it completely phase separates into two domains of type A and type B. But when the same liquid is considered inside a cylindrical pore, after the sudden quench, phase segregation commences with the growth of tiny isotropic domains.  With time, these domains grow and organize themselves into stripes along the axis of the cylinder in a periodic pattern. Therefore, plug-like domains are formed in the absence of wetting interactions between the cylinder wall and the fluid particles~\cite{Liu, Basu}. Finally, the system attains a metastable state and a complete macroscopic phase separation is never achieved. The scenario remains the same far inside the co-existence region also. The width of these domains are found to be insensitive to the length of the cylinder but varies linearly with the pore diameter. 

Nevertheless, when the wetting effect is considered, the growth behavior is quite tangled. In our present study, we analyze the wetting effect on the domain growth dynamics over a wide range of wetting strengths $\epsilon_w$, from partial to full wetting, systematically. The preferential attraction of type A particles implies that the said particles have comparatively lesser surface tension $\gamma_A$ with the wall, than that of the other type $\gamma_B$~\cite{Cahn}. Hence, if $\theta$ is the contact angle between the fluid and wall interface, then according to Young's condition~\cite{Young} $\gamma_{AB} \mathrm{cos}\theta = \gamma_B - \gamma_A$, where $\gamma_{AB}$ is the surface tension between the A and B interface. The condition for partial wetting and complete wetting is deduced from this criterion~\cite{Cahn}. When $\gamma_B - \gamma_A < \gamma_{AB}$, both A and B species are in contact with the surface and the system is only partially wet. On the other hand, when $\gamma_B - \gamma_A > \gamma_{AB}$, the Young's condition is not valid and the B phase is expelled from the wall resulting in the complete wetting of the wall with phase A. 

Following the rapid cooling process, the phase segregation begins as small isotropic domains starts to form inside the pore. The interaction of phase separation and wetting, known as surface-directed spinodal decomposition, involves a dynamic interplay between these two kinetic processes. In Fig.~\ref{fig:ea01} we show the time evolution of the domain structures of our system for the wetting interaction $\epsilon_w = 0.1$. The outcome is more or less close to the phase separation of binary liquid inside the nanopore without wetting~\cite{Basu,Rao}. The system freezes into a multi-domain metastable state and no further domain evolution is observed with time. The reason can be attributed to the following. When the adjacent stripes are separated by more than a characteristic distance, the length scale saturates as a result of a weak contact between the fronts of the neighboring stripes. The plug like structures are formed and the metastable state is stabilized.

\begin{figure}[h]
\centering	
\includegraphics[width=0.48\textwidth]{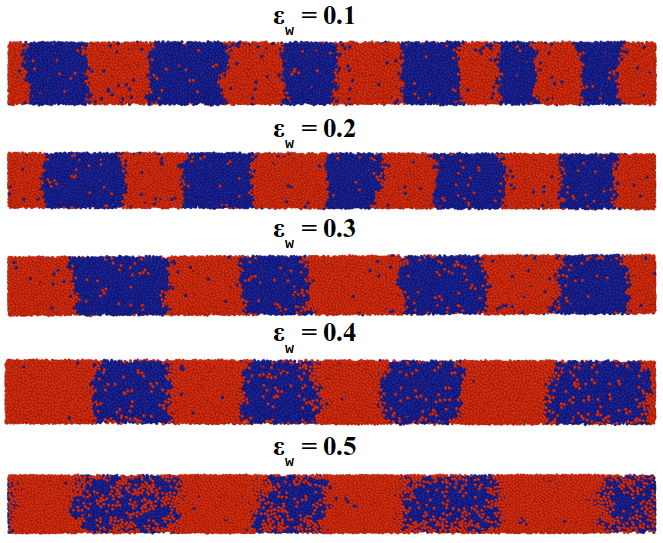}
\caption{Final configurations of our binary liquid system forming plug like domain structure for different wetting interaction $\epsilon_w$.}
\label{fig:final}
\end{figure}

In Fig.~\ref{fig:final} we show the domain structures corresponding to the longest possible simulation time for the wetting interactions $\epsilon_w$ in the range 0.1 to 0.5. It clearly depicts how the metastable state varies with $\epsilon_w$. The width of the striped domains appears to increase as the wetting strength increases. This is because the pore wall acts as a bridge between the alternative stripes which facilitates phase separation with increasing $\epsilon_w$. We find a critical field strength $\epsilon_w = 0.5$ up to which the metastable phase separation takes place with the formation of stripes. Therefore, $\epsilon_w \le 0.5$ corresponds to partial wetting. 

\begin{figure}[h]
	\centering	
	\includegraphics[width=0.45\textwidth]{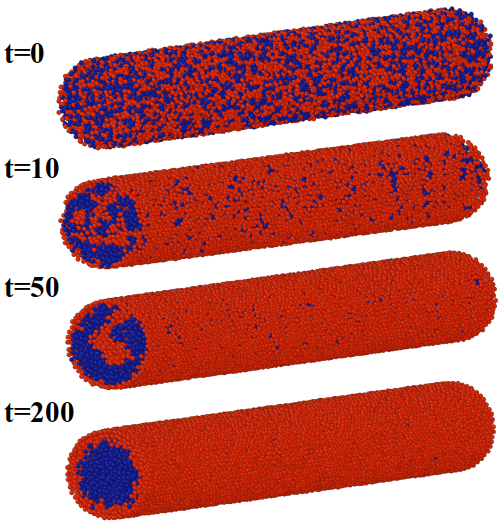}
	\caption{Representative snapshots for the phase separating binary liquid mixture inside the cylindrical pore for the full wetting interaction $\epsilon_w=0.8$.}
	\label{fig:08}
\end{figure}

\begin{figure}[ht]
\centering	
\includegraphics[width=0.4\textwidth]{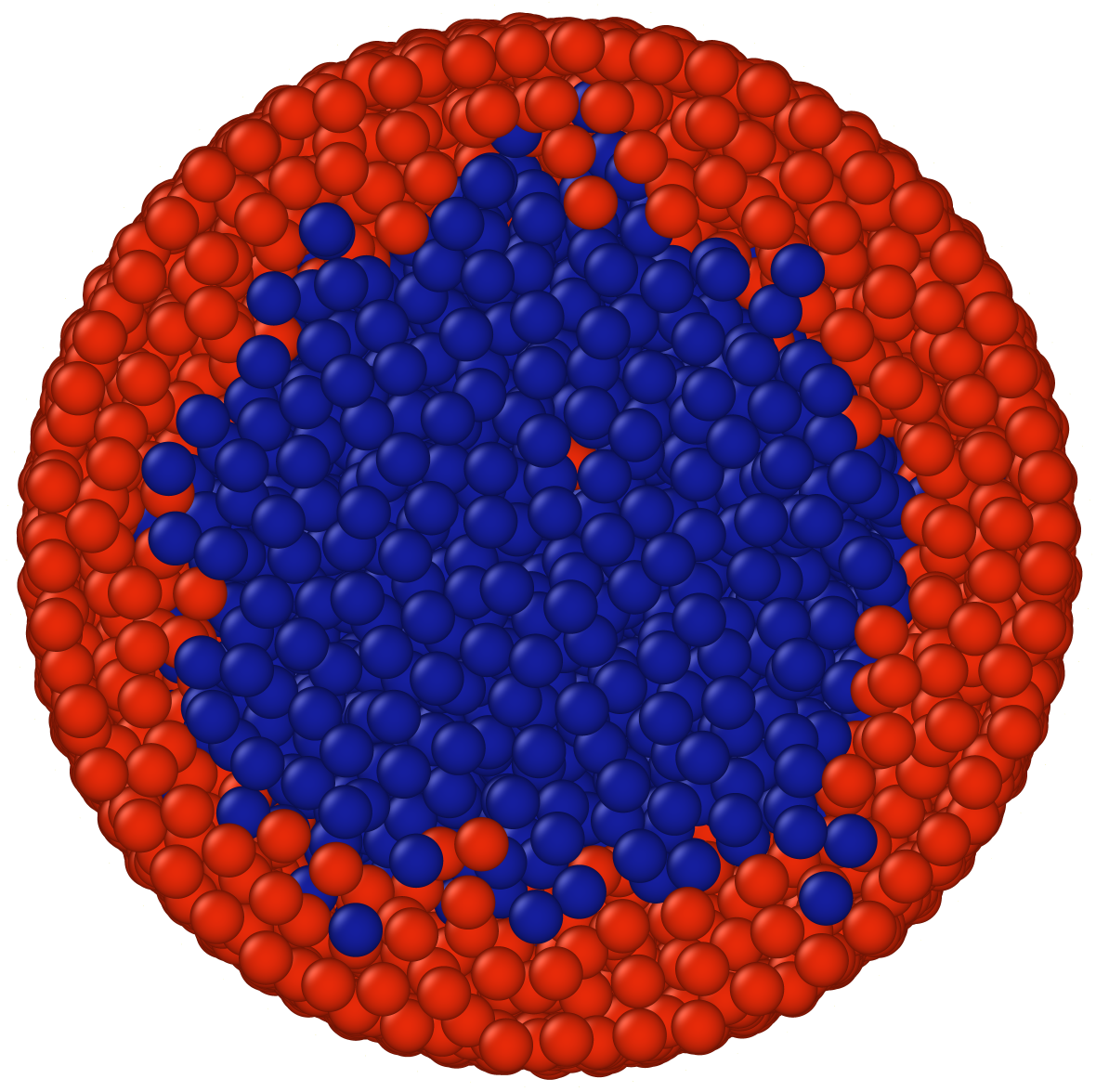}
\caption{The cross sectional view of the fully phase separated liquid inside the pore.}
\label{fig:08cross}
\end{figure}

As the wetting strength is increased further, stripe formation no longer occurs. Instead, the transition from a plug-like to a tube-like domain is observed, which corresponds to the full wetting morphology. In Fig.~\ref{fig:08} we show the time evolution of the domain structure for the highest interaction strength $\epsilon_w = 0.8$ chosen in our simulation. We clearly observe a complete phase separation of the binary liquid inside the pore. For better visualization, the cross sectional view of the system after the complete phase separation is shown in Fig.~\ref{fig:08cross}. This surface field value satisfies the complete wetting condition mentioned above. Correspondingly, the type A particles interact with the pore wall, forming a layer near to it. On the other hand, the neutral B type particles are pushed towards the axis of the cylinder~\cite{Liu}. Thus our simulation confirms that, when the wetting interaction is above a particular threshold value (in our case $\epsilon_w > 0.5$), we find the tube-like domain along the axis of the cylinder formed by the non-wetting particles, whereas the wetting species coats the inner surface of the pore. Therefore, a complete phase separation is achieved for the full wetting case.

Next, we examine the dynamical properties of the system for the partial wetting case, that exhibits stripe formation. Since the geometrical confinement imposed on the system results in the stripe patterned domains, growth is analyzed along the axial direction~\cite{Gelb1,Gelb2}. Therefore, the order parameter takes the form
\begin{equation}
	\label{eq:psi_x}
	\psi(x,t)={\frac{\rho_A(x,t) - \rho_B(x,t)}{\rho_A(x,t) + \rho_B(x,t)} }
\end{equation}
To compute $\psi(x,t)$ we divide the cylinder vertically into sections of equal width $\Delta x=2.0$. The $\rho_A(x,t)$ and $\rho_B(x,t)$ are calculated for each section and thus the $\psi(x,t)$ is obtained from Eq.~\ref{eq:psi_x}. 

To study the domain growth dynamics we compute the two-point equal-time correlation function given by Eq.~\ref{eq:correlation} along the x axis. For the wetting strength $\epsilon_w \le 0.5$, the observation of a consistent self-similarity pattern in stripe formation suggests that our system is likely to adhere to the scaling law $C(x,t) \equiv \tilde C(x/\ell(t))$, where $\tilde C$ is a time independent master scaling function~\cite{Bray}. The identification of this scaling law enables the definition of a time-dependent length scale $\ell(t)$ based on the decay of $C(x, t)$. Throughout the paper, we utilize the first zero-crossing of $C(x, t)$ as a reliable measure of $\ell(t)$ .

\begin{figure}[ht]
	\centering	
	\includegraphics[width=0.45\textwidth]{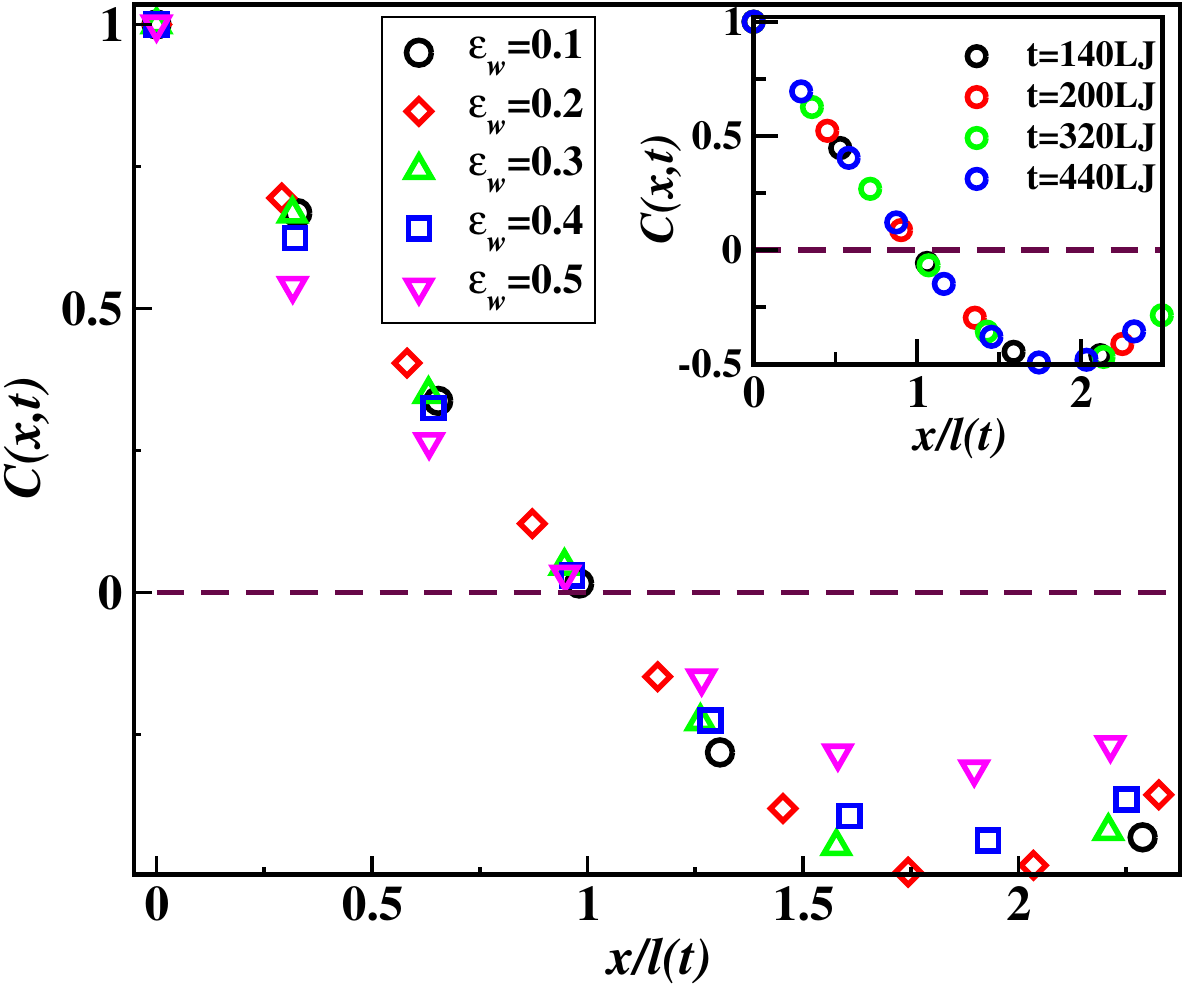}
	\caption{Scaled correlation function $C(x,t)$ vs $x/\ell(t)$ for different partial wetting interaction $\epsilon_w$. In the inset we show the scaling plot of $C(x,t)$ vs $x/\ell(t)$ for $\epsilon_w=0.2$ for different times.}
	\label{fig:stripe correlation}
\end{figure}

Fig.~\ref{fig:stripe correlation} confirms the scaling law of the correlation where $C(x,t)$ is plotted vs $x/\ell(t)$ for different strength of wall interactions corresponding to partial wetting. The data collapse is highly evident, except for the case of $\epsilon_w = 0.5$, where the impact of wetting is close to disrupting the barrier responsible for stripe formation and maintaining a metastable equilibrium. So, we can generalize that Porod law is valid for partial wetting. The inset of the figure shows the scaling of correlation for a particular choice of $\epsilon_w = 0.2$ at different times. We observe an excellent data collapse. A similar scaling behavior is observed for other $\epsilon_w$ values also (not shown here).

\begin{figure}[ht]
	\centering	
	\includegraphics[width=0.45\textwidth]{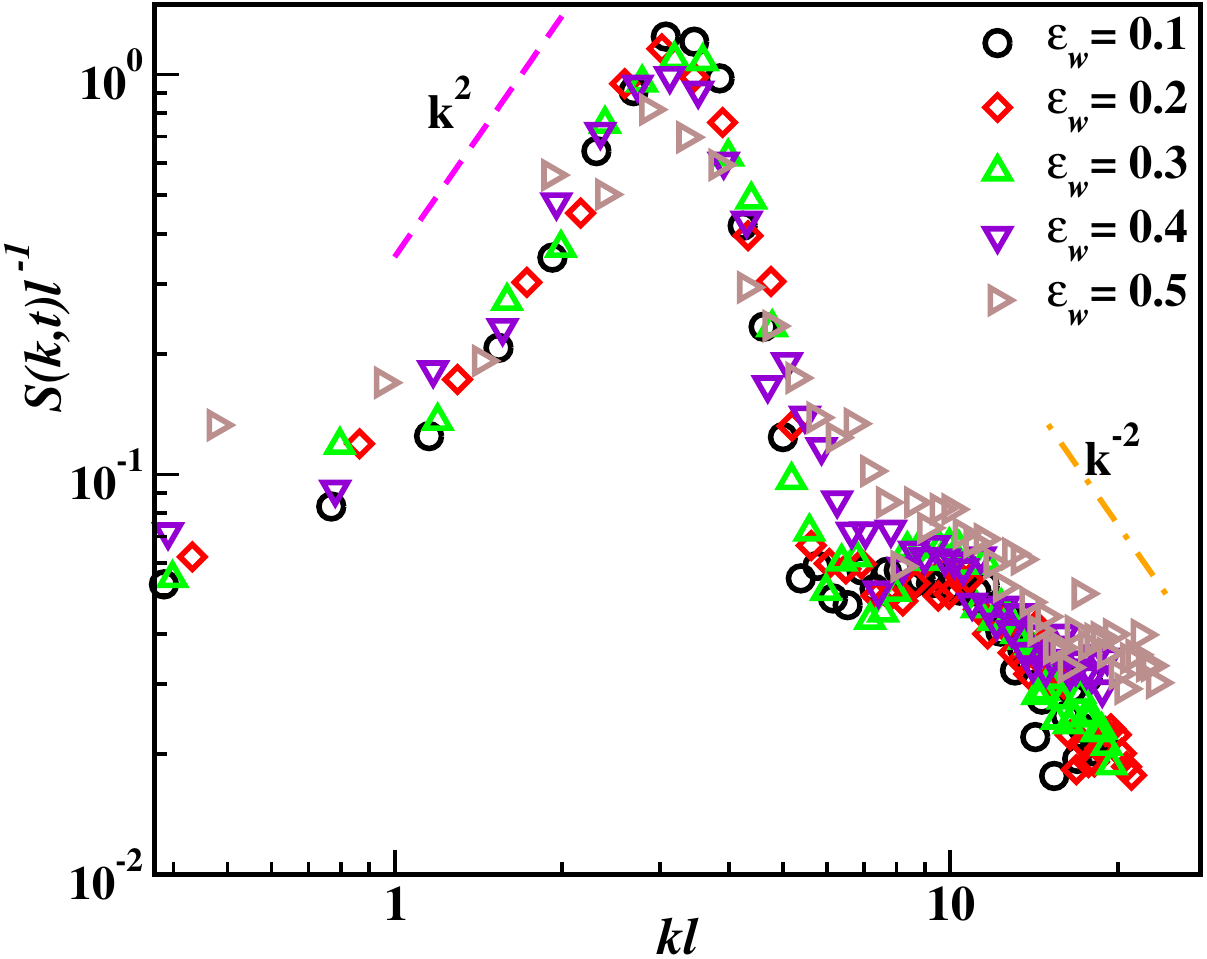}
	\caption{The scaled structure factor $S(k,t)\ell^{-1}$ vs. $k\ell$ for the different partial wetting strength $\epsilon_w$. The dashed lines are the guide line for the Porod law.}
	\label{fig:sk-1d}
\end{figure}

To examine patterns and investigate domain structures in both simulations and experiments, it is a common practice to calculate the structure factor. One-dimensional correspondence of Eq.~\ref{eq:structure factor} is used to calculate the same. Fig.~\ref{fig:sk-1d} shows the scaled structure factor. The decaying part of the tail exhibits a power law $S(k,t) \sim k^{-2}$, showing the Porod law behavior and supporting the one-dimensional growth in the system as defined in Eq.~\ref{eq:porodLaw}. It is also evident that the structure factor associated with $\epsilon_w=0.5$ exhibits a slight deviation from the Porod tail, indicating the critical limit for stripes formation where the interfaces are less distinct and rough. Additionally, the $S(k,t) \sim k^2$ behavior at the small-$k$ limit further supports the argument of one-dimensional domain growth in the system.

\begin{figure}[h]
    \centering
   \includegraphics[width=0.48\textwidth]{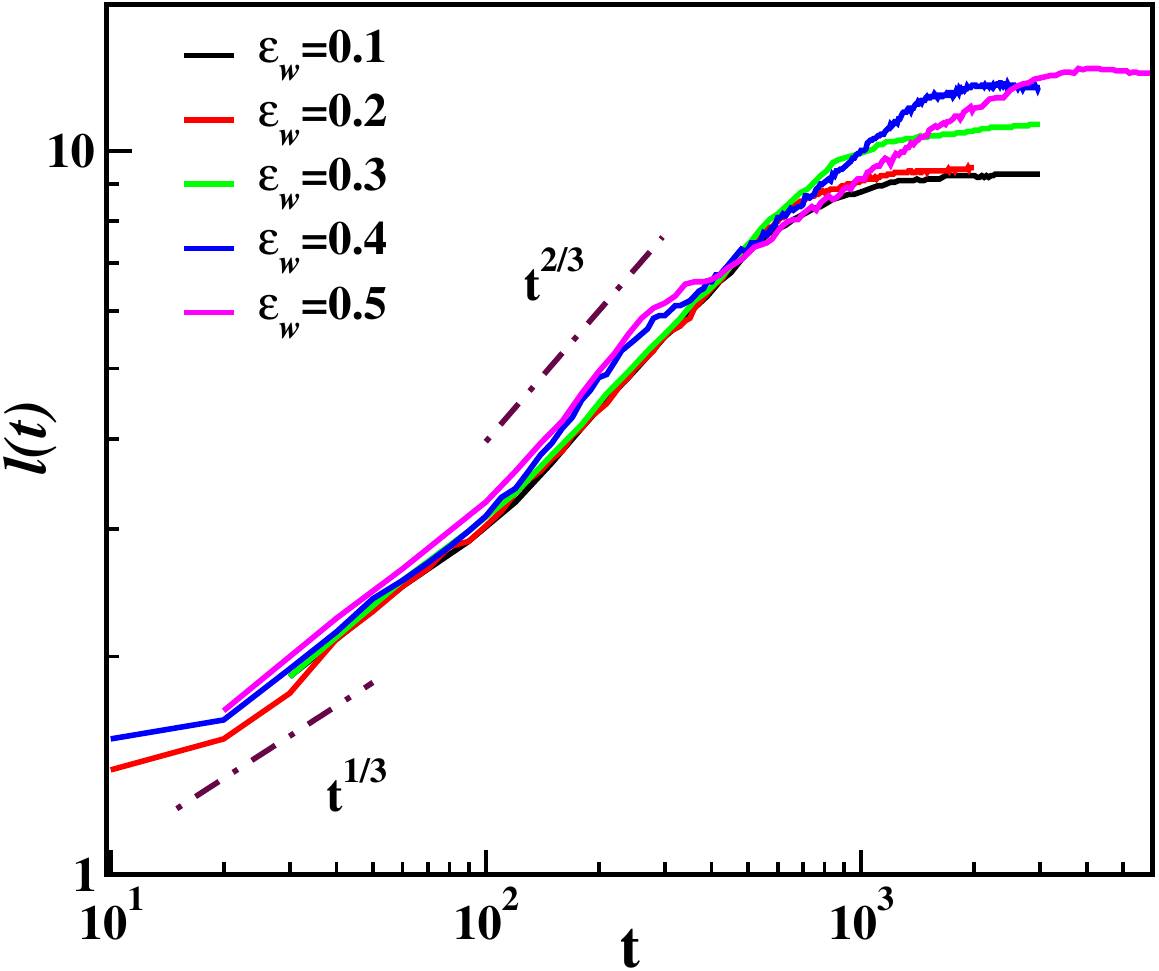}
     \caption{ The time evolution of the length scale $\ell(t)$ for the partial wetting case with different $\epsilon_w$. The dashed lines are the reference for the power law growth.}
     \label{fig:stripe lengthscale}
\end{figure}

The correlation and structure factor clearly show an onset of deviation for $\epsilon_w=0.5$, as they start to diverge from the one-dimensional growth and the wetting effect becomes dominant and tempts to shift towards the capsule-like structure. The pore diameter not being large enough, we do not observe a proper capsule formation. Instead, a direct transition occurs from plug to tube like domains as the $\epsilon_w$ is increased further. A more detailed discussion of this phenomenon will follow. 

Subsequently, our attention turns to quantifying the growth of the stripes along the axial direction in terms of the lengthscale $\ell(t)$. As mentioned earlier, this quantity is computed from the first zero crossing of the correlation function $C(x=\ell,t)=0$. The time evolution of $\ell(t)$ is shown in Fig.~\ref{fig:stripe lengthscale}. The dashed lines in the graph show the power law correspondence at different stages of the growth. The transport mechanism in the system decides the rate of domain growth. During the initial stage, the system exhibits diffusive behavior, adhering to the Lifshitz-Slyozov growth law as $t^{1/3}$. This is followed by a crossover to the inertial hydrodynamic growth characterized by a power law of $t^{2/3}$~\cite{Furukawa1}. The same growth exponents were obtained when the pore wall was considered neutral (no-wetting). It is worth noting that the wetting strength of $\epsilon_w=0.5$ is evidently a critical scenario where, despite the presence of stripe domains, the dynamical properties deviate significantly from the typical behavior observed for partial wetting case.

\begin{figure}[h]
	\centering
	\includegraphics[width=0.48\textwidth]{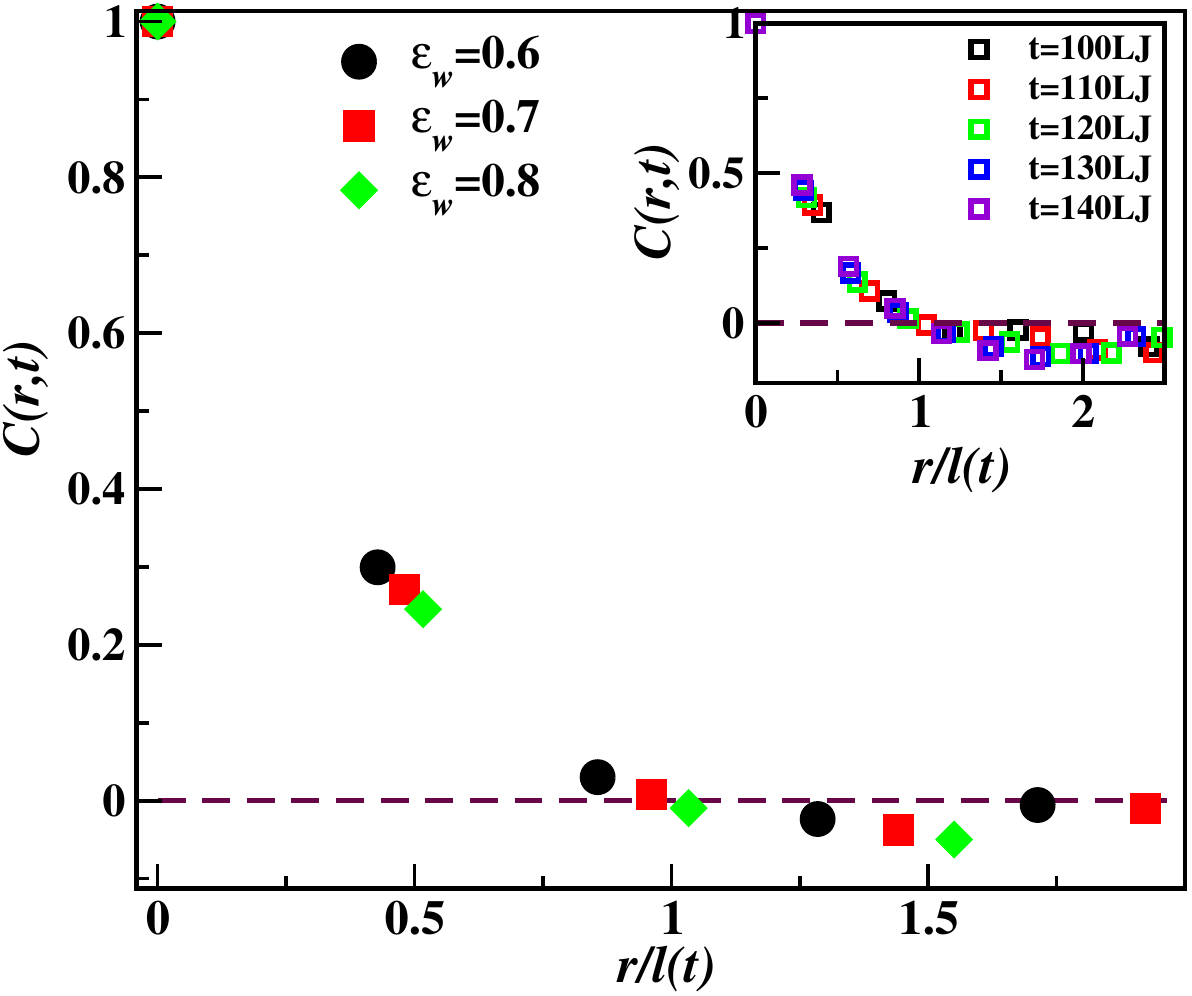}
	\caption{Scaled correlation function $C(r,t)$ vs $r/\ell(t)$ for different $\epsilon_w$ corresponding to the full wetting case. In the inset we show the scaling plot of $C(r,t)$ vs $r/\ell(t)$ for $\epsilon_w=0.8$ for different times.}
	\label{fig:wetting-correlation}
\end{figure}

\begin{figure}[!h]
	\centering
	\begin{subfigure}[h!]{0.9\columnwidth}
		\includegraphics[width=\columnwidth]{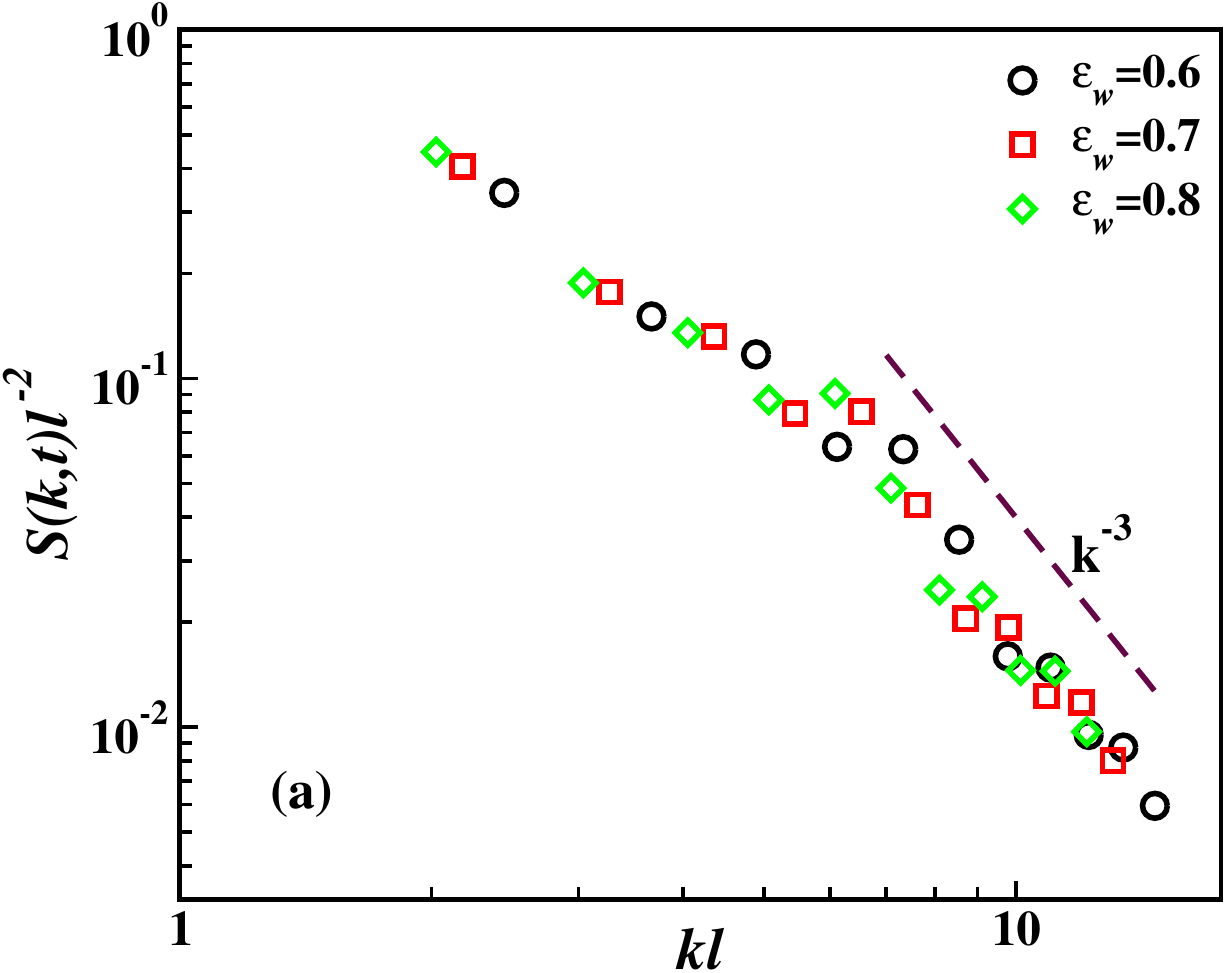}
	\end{subfigure}\\
	\begin{subfigure}[h!]{0.9\columnwidth}
		\centering
		\includegraphics[width=\columnwidth]{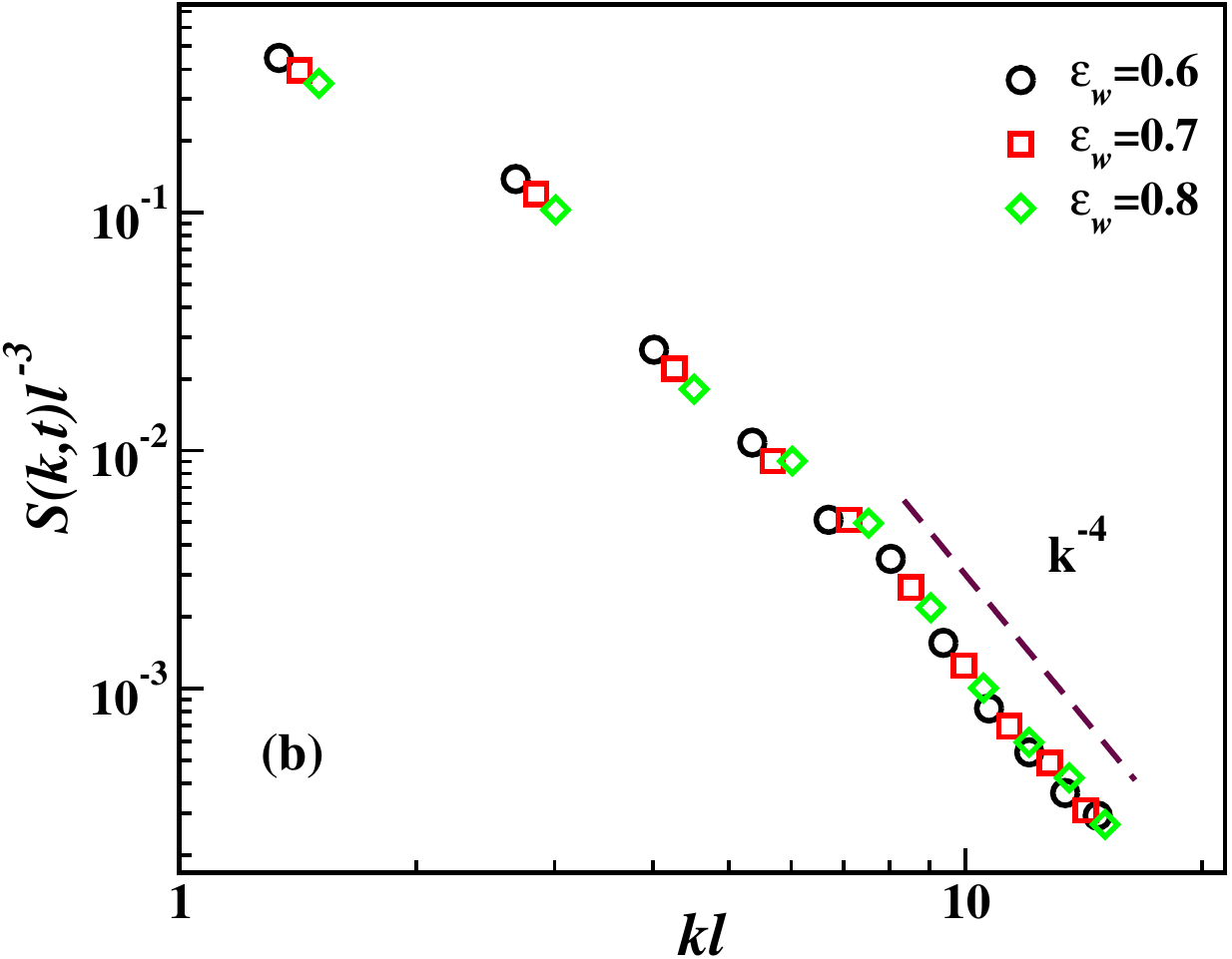}
	\end{subfigure}
	\caption{The scaled structure factor $S(k,t)\ell^{-1}$ vs. $k\ell$ graph for different $\epsilon_w$ corresponding to the full wetting case at a fix time $t=110$ for the (a) wetting species (A particles), (b) non-wetting species (B particles). The dashed lines are the guide line for the Porod law.}
	\label{fig:sk-2d}
\end{figure}

Next, we shift our attention to investigate the phase separation dynamics for the full wetting case, i.e. $\epsilon_w>0.5$. A complete phase separation is achieved here via the formation of tube-like domains. A typical  domain morphology is displayed in Fig.~\ref{fig:08}. It is crucial to emphasize that during the phase separation of domains inside the pore for the complete wetting case, the correlation is assessed radially rather than axially. Hence the order parameter is calculated accordingly from Eq.~\ref{eq:order_parameter}. For that, the whole system is divided into small cubic boxes of size $(2\sigma)^3$ and the local density fluctuations are computed over these boxes. Finally, we calculate the correlation function along the radial direction from Eq.~\ref{eq:correlation}. 

During the coarsening process, the two species proceed individually following the surface directed spinodal decomposition. The wetting species endures surface enrichment while the other is expelled from the surface. This results in complete phase separation, as shown in Fig.~\ref{fig:08}.  The correlation function for both the species is calculated separately to study their individual domain growth. Fig.~\ref{fig:wetting-correlation} corresponds to the scaled correlation of wetting particles. We observe a satisfactory data collapse for different interaction strength $\epsilon_w$. The inset shows the scaled correlation for the maximum interaction strength $\epsilon_w=0.8$ at different times. They exhibit a perfect data collapse as well. Hence, the surface directed migration of particles in our confined system perseveres and upholds the presence of superuniversality and Porod law~\cite{Binder2,Lai,Corberi}. The same exercise is repeated for the non-wetting species (not shown), and a similar scaling behavior is observed.

Considering the rationale mentioned earlier, it is prudent to compute the structure factor independently for each of the species. The results are shown in Fig.~\ref{fig:sk-2d} for three different $\epsilon_w$ at a particular time $t=110$. The dotted lines correspond to the power law reference. The results clearly demonstrate that the trailing section of the structure factor exhibits distinct power laws for the two species. According to Eq.~\ref{eq:structure factor}, wherein $d$ represents the dimension of the domain, $k^{-3}$ pertains to growth in two-dimensions while $k^{-4}$ refers to growth in three-dimensions. This suggests that the wetting species experiences two-dimensional domain growth, while the other undergoes three-dimensional growth. This can be clearly understood from Figs.~\ref{fig:08} and \ref{fig:08cross}, where the type A particles form a layer on the inner surface of the pore wall, resembling a curved two-dimensional plane. Therefore, the structure obtained for the wetting particles is two-dimensional, providing a rationale for the Porod law exponent. On the other hand, type B particles that congregate around the axis of the cylindrical pore behave akin to a bulk system. This three-dimensional structure of the non-wetting particles is affirmed by the Porod tail behavior observed in the structure factor.

\begin{figure}[h]
	\centering
	\includegraphics[width=0.5\textwidth]{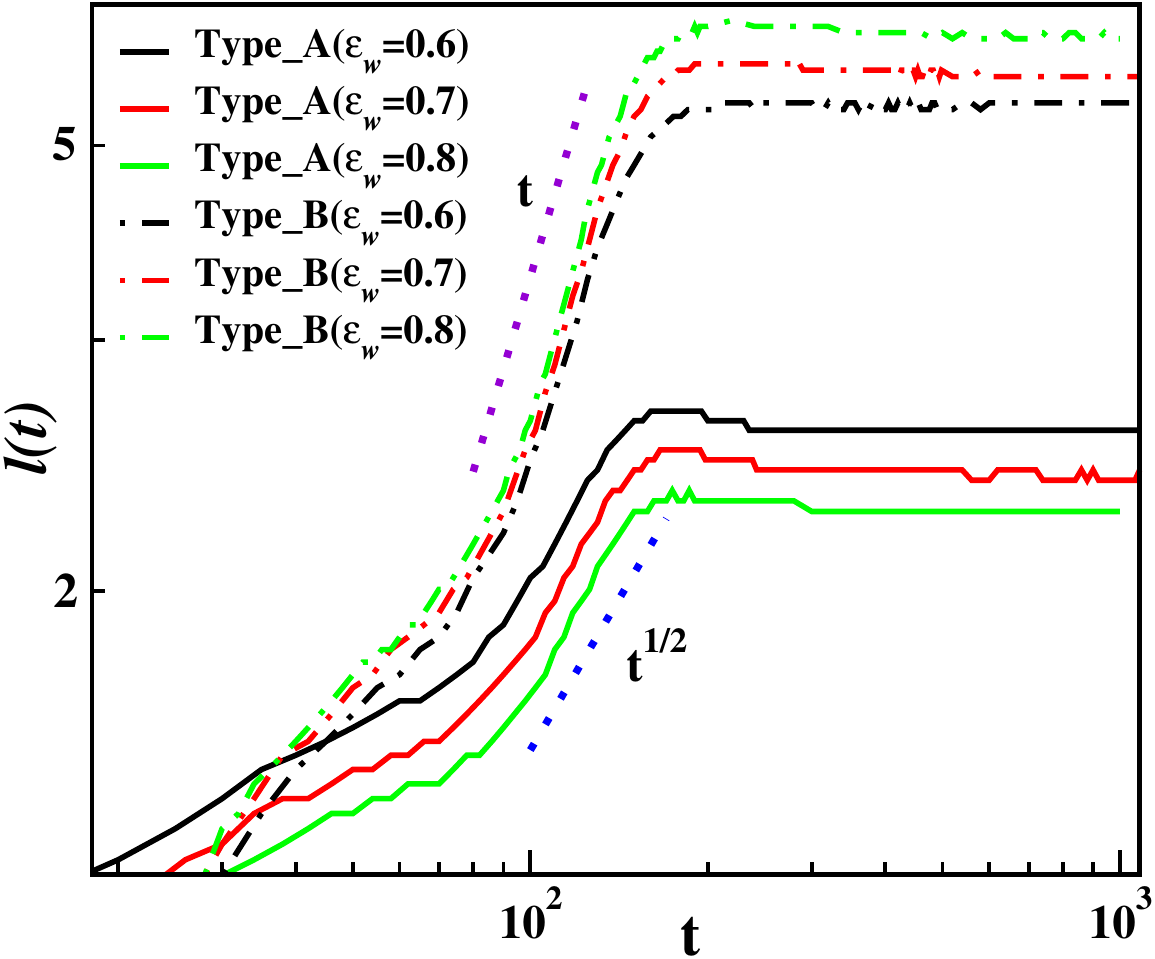}
	\caption{The time evolution of the length scale $\ell(t)$ for the full wetting case with different $\epsilon_w$. The solid lines and dashed dotted lines refer to the wetting and non-wetting species respectively. The dotted lines represent the power law growth.}
	\label{fig:wetting lengthscale}
\end{figure}

The time-dependence of the characteristic domain growth for the two types of particles are computed separately for three different $\epsilon_w$. In Fig.~\ref{fig:wetting lengthscale} we show the $\ell(t)$ for both the species. The dotted lines indicate the power law. The domain growth of  the non-wetting species resembles liquids in three-dimensional bulk system. After an initial transition period, the domain size grows as $\ell(t) \sim t$, which corresponds to the bulk viscous hydrodynamics growth. This result is consistent with the structure factor which shows a three-dimensional Porod tail of $k^{-4}$.

For the wetting species, the growth law is found to be $t^{1/2}$, which resembles domain growth of liquids in two-dimensional surface. This can be comprehended as follow. The wetting particles interact with the pore wall and form layers inside the wall which specifically encloses the non-wetting particles. Therefore, this structure is identical to the two-dimensional curved surface. It is well known that a binary liquid phase separates with a growth law exponent of $1/2$ on a two-dimensional plane. Hence the domain growth of the wetting particles can be explained analogously. This is further endorsed by the structure factor in Fig.~\ref{fig:sk-2d}, which shows Porod tail behavior of $k^{-3}$.

\section{Conclusion}
In summary, we have studied the surface directed spinodal decomposition of a segregating binary liquid mixture system confined inside cylindrical pore using comprehensive molecular dynamics simulation. One of the species of the liquid adheres to the pore surface, whereas the other remains inert. A wide range of wetting interactions is being contemplated, encompassing both partial and full wetting. For the partial wetting case, the domain structure resembles no wetting scenario. After the initial domain growth, phase separation is halted via formation of plug-like structures. The growth exponent of the domain is estimated to be 2/3 which suggests an one-dimensional growth dynamics. This is further confirmed from the Porod law tail of the structure factor. 

The scenario changes completely as the wetting interaction is increased beyond a critical value ($\epsilon_w>0.5$). The plug-like structure breaks down and cylindrical domains emerge for the full wetting case. Hence, a complete phase segregation is observed when the wetting substance migrates toward the pore surface and creates layers that encompass the non-wetting species located around the axis of the cylinder. The wetting substance is observed to adhere to a two-dimensional domain growth pattern, characterized by the growth exponent $\alpha=1/2$ in the viscous hydrodynamic regime. This is supported by the Porod tail pertaining to the structure factor. On the other hand, the non-wetting species is found to experience linear domain growth over time. This implies that the non-wetting species behaves similarly to a three-dimensional bulk system. This behavior was additionally affirmed through an examination of the tail section of the structure factor. Our works provides a comprehensive understanding of the kinetics of phase separation in confined liquids under different wetting conditions. It will be interesting to extend this work where the confinement has complex topology, i.e. random porous media~\cite{Rounak2}. \\

\noindent{\it Acknowledgement.---} B. Sen Gupta acknowledges Science and Engineering Research Board (SERB), Department of Science and Technology (DST), Government of India (no. CRG/2022/009343) for financial support. Daniya Davis acknowledges VIT for doctoral fellowship.

\end{document}